\documentclass{aa}
\usepackage{graphicx}
\begin{document}
%\thesaurus{06 (19.53.1; 19.63.1)} 

\title{Comparative seismology of pre- and  main
sequence stars in the instability strip}

\author{ Marian Suran$^1$, MarieJo Goupil$^2$, Annie
Baglin$^3$,\\  Yveline Lebreton$^2$, Claude Catala$^4$}
\institute{$^1$ Astronomical Institute of Romanian Academy,\\
$^2$ DASGAL, UMR CNRS 8633, Observatoire de Paris-Meudon,\\
$^3$ DESPA, UMR CNRS 8632, Observatoire de Paris-Meudon,\\
$^4$LAT, UMR CNRS 5572, Observatoire de Midi-Pyr\'en\'ees }

\offprints{M. Suran, \email{suran@roastro.astro.ro}  }

\date{}

\abstract{
Pulsational properties of 
 1.8 M$_{\odot }$  stellar models 
covering the latest stages of contraction toward the main
sequence up  to early  hydrogen burning phases 
are investigated by means of linear nonadiabatic analyses.
Results  confirm that pre-main sequence stars (pms) which cross the classical 
instability strip on their way toward the main sequence are 
 pulsationally unstable with respect to the
classical opacity  mechanisms.
For both pms and main sequence  types of models  in the lower part of the instability 
strip, the unstable frequency range is found to be roughly the
same. Some  non-radial unstable modes are very
sensitive to the deep internal structure of the star.  
It is shown that discrimination between 
pms and main sequence  stages 
is possible using differences in their oscillation frequency distributions
in the low frequency range.
}

\titlerunning{Comparative seismology of pre- and  main
sequence stars}
\authorrunning{Suran et al.}
\maketitle

\section{Introduction}

In the vicinity of
 the main sequence, the HR diagram is confusing as stars of similar
global properties but different stages of evolution lie at the same position.
 In some cases the young pre-main sequence stars
are recognized through specific characteristics, for instance the presence of
nebulosity or high degree of activity. But, this is not always the case.
An  alternative is 
 to take advantage of seismological pieces of 
information whenever it is possible. 
We therefore  carry out  a
comparative study of the pulsational behavior of pre-main sequence (hereafter 
pms) and  main sequence (hereafter postZams)  stars of intermediate mass, 
in the classical instability strip.

 Our study is restricted to the
latest stages of the pre-main sequence contraction when traces of the
formation phases have already disappeared, so that a pms model can be
built with the quasi-static approximation.

The outer layers of pms and postZams stars
 having almost the same effective temperature and gravity are very
similar. As these layers drive the pulsation in this range of temperature,
it is reasonable to expect that pms stars in the instability strip are
also destabilized by classical opacity mechanisms and that the same type of
 modes  than for  the postZams stars are excited. 
This idea has been confirmed by 
 Suran (\cite{sur}) and  in a more detailed work but for pure 
radial modes of very young and cool stars
by  Marconi \& Palla  (\cite{mp}). This supports  the   vibrational
instability as an explanation for the observed periodic variations
 detected for a  few stars  which are  
suspected to be in the pre-main
sequence stage 
(Kurtz \& Marang \cite{km}, Catala {\sl et al.} \cite{cat}, Kurtz, 1999).

Differences, on the other hand,  exist between the two stages: whereas pms has
some relicts of its gravitational contraction phase, the main sequence
nuclear burning modifies the inner core structure, and postZams models
 develop chemical inhomogeneities. 
In a preliminary work, Suran (\cite{sur}) has stressed some 
consequences of these differences in the structures 
of the frequency distributions. The present 
 paper  further investigates and compares 
the pulsational properties of these two phases for non-radial oscillations.
It is organised as follows: stellar modelling and its  physical inputs are
described in Sect. 2, where comparable pms and postZams models are
selected. Oscillation frequencies associated with nonradial 
 modes for both types of objects are
compared in Sect. 3 and 4, and emphasis is put on the structure of the 
frequency spectra in both stages.
 Finally in Sect. 5, we discuss the possibility to
infer the evolutionary stage, i.e. the age of a variable object in this
region of the HR diagram, through the pattern of its frequency spectrum.

\section{ Static models and evolutionary sequences }

\subsection{ Evolutionary tracks }

In the early stages of pms evolution, various 
rapid- dynamical and thermal- processes occur 
which modify the internal structure of the protostar.   
By the time the protostar reaches the classical instability strip 
on its way toward the main sequence, these complex phenomena 
have long disappeared  and the evolution has considerably slowed down.
The usual quasi-static approximation then holds and is  assumed here. 
Rotation is not considered and the star is modelled  in 
a spherically symmetric equilibrium state.

Equilibrium models have been computed 
with  CESAM  (Morel \cite{mo}), 
an evolutionary code with a numerical accuracy of first order in time 
and third order in space. The models are built with  
 the fast version of the code i.e. 300 mesh points in mass;
 pulsation calculations on the other hand use an
extended 2000 mesh grid, calculated  from  the static model by means of  the
basis of the spline solution.

The EFF equation of state (Eggleton {\sl et al.} \cite{eff})
is used as it  is sufficient for our purpose
in this mass range. 
Opacities are provided by the  Livermore Library
 (Iglesias \& Rogers \cite{ir}), complemented at 
low temperatures ($T \le 10000K$)  by 
Alexander \& Ferguson 's (\cite{AF}) tables. 
The nuclear network  contains the following species 
$^1$H, $^3$He, $^4$He, $^{12}$C, $^{13}$C, 
$^{14}$N, $^{15}$N, $^{16}$O, $^{17}$O.
The main  nuclear reactions of pp+CNO cycles are included 
with the species $^2$H, $^7$Li, $^7$Be set  at equilibrium.
Nuclear reaction rates
are taken from Caughlan \& Fowler (\cite{ca}); weak screening is assumed.
The  isotopic helium ratio is fixed at
$He^3/He^4=1.01\,10^{-4}$ and 
the mixture of heavy elements is solar
(Grevesse \& Noels \cite{gn}).
In the convection zones, the temperature gradient is
computed according to the standard mixing-length theory, with
the mixing-length defined as $l\equiv \alpha H_{\rm p}$,
where $H_{\rm p}$ is the pressure scale height. Assuming a constant
value of $\alpha$ along the main sequence (Fernandes {\sl et al.} \cite {fer}), we have
chosen
$\alpha =1.67~$, the value obtained  for a calibrated Sun
computed with the same physical assumptions.
The atmospheric layers are treated with the Eddington's grey approximation.

We consider the lower part of the instability strip where one
 encounters the group of $\delta $ Scuti stars defined here as Pop I stars 
in a  postZams stage. 
The associated mass range  approximately is 1.6-2 M$_{\odot }$.
 A mass of  1.80 M$_{\odot }$ with a
standard chemical composition for Population I (X=0.71, Y=0.27, Z= 0.02)
 is therefore chosen as typical for the present work.

Fig.1 shows the evolutionary track running from the latest pms stages to the
early postZams ones for this  1.80 M$_{\odot }$ stellar model.
Peculiar stages are indicated  and their characteristics listed in Table 1.
Local maxima of the luminosity  along the pms track
 correspond  to the onset of rapid nuclear  burning stages: 
ignition of the p-p reaction occurs close to models M6 and M7; 
the C$^{12}$-N$^{14}$ transformation (close to models M12, M13 and M14)
 gives rise to core convection.
 As the CNO
cycle contribution becomes increasingly important and eventually dominant, 
the convective core slightly expands till the stage of
M10.  Once the M14 stage is reached, 
the star definitely settles into the
CNO cycle burning phase.
The zero age main sequence (Zams) model is defined as
the model with the minimum luminosity  just 
before evolution starts back toward the red. At this stage, M15 in Table 1, 
the gravitational energy ($\epsilon_{grav}$) 
 represents less than 1\% of the total generated  energy ($\epsilon_{tot}$)
and the central initial hydrogen  content  has only been slightly 
modified (see Table 1). 

This study does not include the earliest stages, hence 
it  cannot provide exact
values of ages. However, evolutionary time scales for phases 
after the ignition of  
the central hydrogen burning phase are reliable. 
We have therefore arbitrarily chosen the stage M1 as the initial epoch.

\begin{figure}
\includegraphics[width=8.5cm]{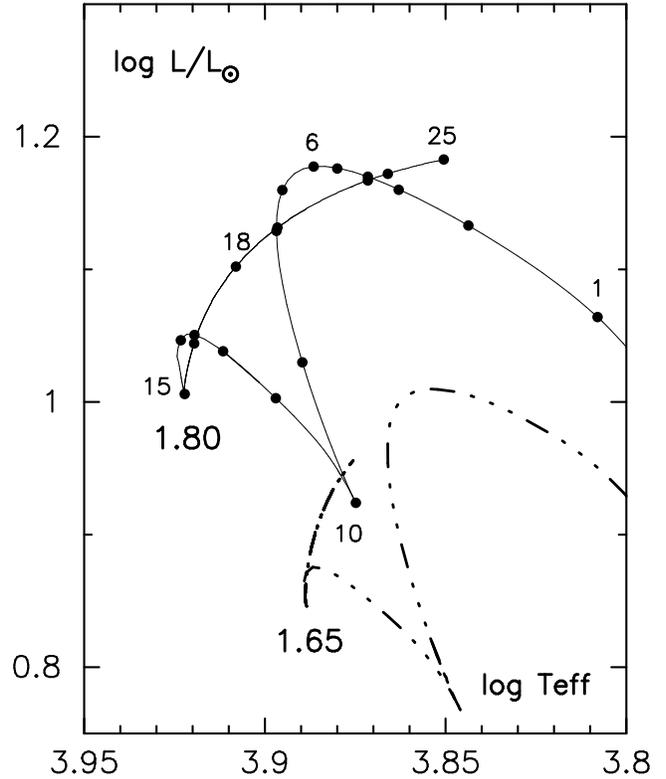}
\caption[]{Evolutionary sequence of a 1.80 M$_{\odot }$ model
(solid line). Selected models listed in Table 1 are indicated by
dots and labeled on the track. A 1.65 M$_{\odot }$ evolutionary
sequence is also shown for comparison (dashed line).}
\end{figure}

\begin{table}
\caption[]{Selected stellar models from the  1.80 M$_{\odot }$ evolutionary
sequence. The second column gives the effective temperature $T$ of the model.
 Ages are in Myr, luminosities L and total stellar radii,$R$, in solar
units,  convective radii $r_{conv}$in unit of total stellar radii. $X_c$ is the
central hydrogen content in percentage}
%\label{aa}
\begin{center}
\begin{tabular}{|l|l|l|l|l|l|l|} \hline
 &  log T & log L & R & age & $X_c$ &
$r_{conv}$\\
 \hline
 M 1  &  3.808 &  1.064 &   2.751 &        .0 &   0.7143 &   0.000  \\
 M 2  &  3.844 &  1.133 &   2.527 &        .4 &   0.7143 &   0.000  \\
 M 3  &  3.863 &  1.160 &   2.389 &        .6 &   0.7142 &   0.000  \\
 M 4  &  3.872 &  1.170 &   2.319 &        .7 &   0.7142 &   0.068  \\
 M 5  &  3.880 &  1.176 &   2.251 &        .8 &   0.7142 &   0.124  \\
 M 6  &  3.887 &  1.177 &   2.184 &        .9 &   0.7142 &   0.166  \\
 M 7  &  3.895 &  1.160 &   2.056 &        1.1 &   0.7142 &   0.221  \\
 M 8  &  3.897 &  1.129 &   1.970 &       1.25 &   0.7142 &   0.243  \\
 M 9  &  3.890 &  1.030 &   1.816 &        1.6 &   0.7141 &   0.258  \\
 M10  &  3.875 &  0.924 &   1.721 &        2.4 &   0.7139 &   0.239  \\
 M11  &  3.897 &  1.003 &   1.702 &       3.8 &   0.7136 &   0.211  \\
 M12  &  3.912 &  1.038 &   1.657 &       4.4 &   0.7135 &   0.186  \\
 M13  &  3.919 &  1.050 &   1.621 &       4.8 &   0.7134 &   0.186  \\
 M14  &  3.923 &  1.047 &   1.586 &       5.2 &   0.7134 &   0.201  \\
{\bf M15 } & {\bf 3.922}& {\bf 1.006}& {\bf 1.521}  & {\bf 11.}&  {\bf 0.7120}
& {\bf 0.205}  \\
 M16  &  3.920 &  1.044 &   1.608 &      300. &   0.6383 &   0.204  \\
 M17  &  3.918 &  1.058 &   1.649 &      400. &   0.6097 &   0.202  \\
 M18  &  3.908 &  1.102 &   1.813 &      700. &   0.5127 &   0.197  \\
 M19  &  3.897 &  1.131 &   1.978 &      900. &   0.4355 &   0.191  \\
 M20  &  3.893 &  1.139 &   2.029 &      950. &   0.4142 &   0.190  \\
 M21  &  3.889 &  1.146 &   2.086 &     1000. &   0.3920 &   0.188  \\
 M22  &  3.876 &  1.162 &   2.248 &     1120. &   0.3340 &   0.182  \\
 M23  &  3.872 &  1.167 &   2.312 &     1160. &   0.3132 &   0.180  \\
 M24  &  3.866 &  1.172 &   2.383 &     1200  &   0.2913 &   0.178  \\
 M25  &  3.851 &  1.183 &   2.594 &     1300. &   0.2320 &   0.170  \\
\hline
\end{tabular}
\end{center}
\end{table}

\subsection{ Structural differences between  pms ans postZams models with
 same global properties}

As mentionned earlier, a source of ambiguity comes from
 the fact that pms and postZams stars  have 
similar surface properties and therefore 
lie at the same position in a HR diagram. 
For instance, our $ 1.8 M_\odot$ sequence in Fig.1 shows that 
the  three pms models   M4,  M8, M13 can be confused with
 the three postZams models M23, M19, M16 respectively, on 
the basis of their sole location in the HR diagram (Table 1). 
The question we address here is
whether small differences in the structures of the pms and 
postZams  models  can induce significant 
changes in their oscillation spectra.
The second  issue is to determine whether these
 frequency differences can be used to discriminate between the two evolutionary
 stages. 

Accordingly, our procedure  consists in 
comparing  seismological properties 
of  pms and postZams models having the same  position in the HR
 diagram and  the same mass (herafter `associated models').  

As expected, the outer layers of associated models 
are very similar: density and temperature profiles and therefore 
the opacity profile are almost identical, we therefore refrain
from showing them.

On the other hand, the central regions significantly  differ.
A  pms model is still slightly contracting 
 and remains chemically homogeneous. 
In contrast,  a postZams model has a
core which traces its nuclear history, changes 
its chemical composition and density, and builds regions with $\mu$
gradients.

Our youngest models
correspond to the phase of the beginning of the $p-p$ chain, with a very small expanding
convective core. During the main sequence phase, as the star ages,
hydrogen is depleted in the central region
and the fully mixed convective core receeds with time starting at
the Zams stage (approximately M15). 
\begin{figure}
\includegraphics[width=8.5cm]{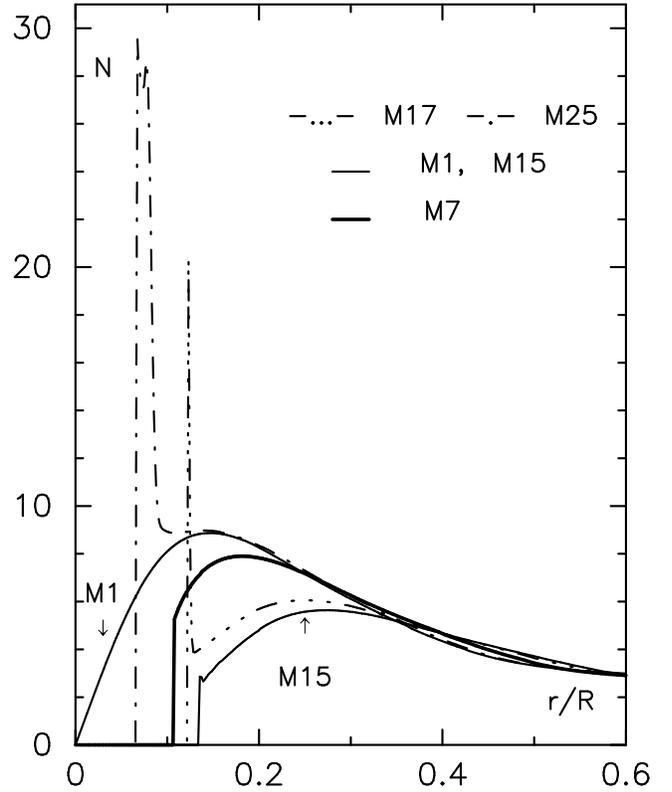}
\caption[]{ Behavior of the Br\"unt-Vaiss\"al\"a frequency $N$ 
in normalized unit (see Eq.1) as a
function of radius (normalized to the total radius) for models 
 along the 1.80 M$_{\odot }$ evolutionary sequence:   
$N$ profiles for  two pms  models (M1, M7),  
Zams model (M15) and two postZams
models (M17 and M25) are represented.}
\end{figure}

Signatures of these modifications are  
clearly  seen on the
profile  of the Br\"unt Vaiss\"al\"a frequency, $ N$, as a
function of radius (Fig.2). The sharp transition between zero  values of $N$ 
within the convective core and positive values 
in the overlaying radiative layers delimits the edge of the convective 
core. As it can be seen in Fig.2, during the pms evolution, the outward 
 progression of the edge of  the expanding convective core 
can be followed with the correlated outward progression of the
inner edge of $N$. The inner $N$ maximum also moves 
outward and decreases.  When the model 
evolves beyond the Zams on the main sequence, 
the inner edge and maximum of the Br\"unt-Vaiss\"al\"a
frequency   receed inwards, while a narrower but
higher peak   associated with the $\mu$ gradient  region  
forms. It can be expected from Fig.2 that modes with normalized frequencies
(Eq.1 below) in the range 4-10  are the most sensitive to the size of the 
convective core and consequently to the evolutionary status (pms or postZams) 
of the star.

\section{ Nonradial pulsations of pms and postZams models}

At every stage considered here, the time scale  of evolution is much 
longer than the pulsation period.
 For instance our $1.8 M_\odot$ sequence indicates time intervals of about 
$ 0.1- 1$  Myr between specific stages of evolution represented 
 in Fig.1. Periods of oscillation of $\delta$ Scuti stars typically range 
between 1h- 3h, reaching roughly  20 h for unstable $g$-modes. 
Growth rates are of the order 
of 100-500 yr, still much smaller than  evolutionary time scales.
The oscillation
amplitudes are expected to grow or decay on a much faster (Kelvin-Helmoltz) time scale
than the (nuclear) evolutionary
time-scale, so that the amplitudes have settled onto their
saturating value (limit cycle) long before evolution can have any disturbing
influence.
 These considerations  authorize investigations  of the pulsations
of pms stars  by means of  usual linear nonadiabatic analyses  of static models 
at a given stage of evolution (Cox \cite{cox}, Unno {\sl et al.}, \cite{un}).
We treat the nonadiabatic effects  in the  formulation of
 Unno {\sl et al.} (1989), using Suran's oscillation code.
The oscillation eigenfrequencies (rad/s) 
are denoted as $\omega$. We will also use 
frequencies in $\mu$Hz, $\nu$,  and normalized frequencies, $\sigma$
defined as 
\begin{equation}
\sigma = {\omega \over  \sqrt{GM/R^3}} 
\end{equation}

Frequencies of the modes are labelled in a usual way: $n,l,m$ for the
radial order, degree and azimutal number. The radial order of a given mode is
determined according to Christensen-Dalsgaard \& 
Berthomieu (1992) with $n=1$ for  the fundamental radial mode. 
The growth rate is obtained directly from
numerical computations but we rather consider   the dimensionless 
$\eta$ parameter (Stellingwerf, 1978; Dziembowski (\cite{dz}):)
\begin{equation}
\eta =\frac{W}{\int_0^1  {\displaystyle \vert w(x) \vert } dx} 
\end{equation}
 where $x=r/R$ and $W$ is  the total mechanical
work  (Cox, 1980; Saio \& Cox, 1980)
\begin{equation}
W = \int_0^1 w(x)  dx = {{\pi \over  E_k} \int_0^M  {P\over \rho}
  Im\lbrace \bigl( {\delta P \over P} \bigr)^*
 \bigl( {\delta \rho \over \rho}\bigr)\rbrace dm} 
\end{equation}
where $\delta P, \delta \rho$ are  the Lagrangean variations of 
the pressure and density respectively. 
The work integral, $W$,  is normalized with the total
 kinetic oscillation energy:
\begin{equation}
E_k = {1\over 2} \omega^2 \int_0^1 \vert {{\bf \delta r}}   \vert^2 \rho ~{\bf d^3r}
\end{equation}
where $\delta r$ is the Lagrangean radius variation.

Interaction between pulsation and convection is   neglected. 
This is certainly  not justified for the  cool stars but the effect is less
important for stars close to the Zams. 
Neglecting this  interaction equivalently affects
 pms and postZams models at the same effective temperature close to the Zams.
Hence this cannot  significantly affect the
main conclusions of the  present comparative study.

\subsection{Evolution of the power spectrum with stellar age}

Fig.3 displays  the  evolution of  the normalized frequencies, $\sigma$,
for a 1.8 $M_\odot$  model evolving from stage 1 (pms) 
to stage 25 (postZams) along the evolutionary track  shown in  Fig.1.
 The plot is restricted to 
$\ell= 0-2$ modes. These modes are usually thought to be
 the easiest modes to be detected, as their visibility
coefficients remain large after integration over the whole stellar disk.
The radial order interval is chosen such as to cover  the range of 
vibrationally unstable modes.
Evolution is represented  in abscissae  by a quantity which is
defined so as to  change monotonously
throughout the pms and main sequence stages: $\Delta R$ is the difference  
between the radius of the model, $R$, 
 and the radius of a reference model, $R_{eff}$, which for
obvious reasons is chosen to be the Zams model (M15 model of Table 1). 
Time goes from left to right. For pms models, 
$\Delta R =R-R_{ref} \leq 0 $ and  for postZams models 
$\Delta R =R_{ref}-R \geq 0$.

\begin{figure} 
\includegraphics[width=8.5cm]{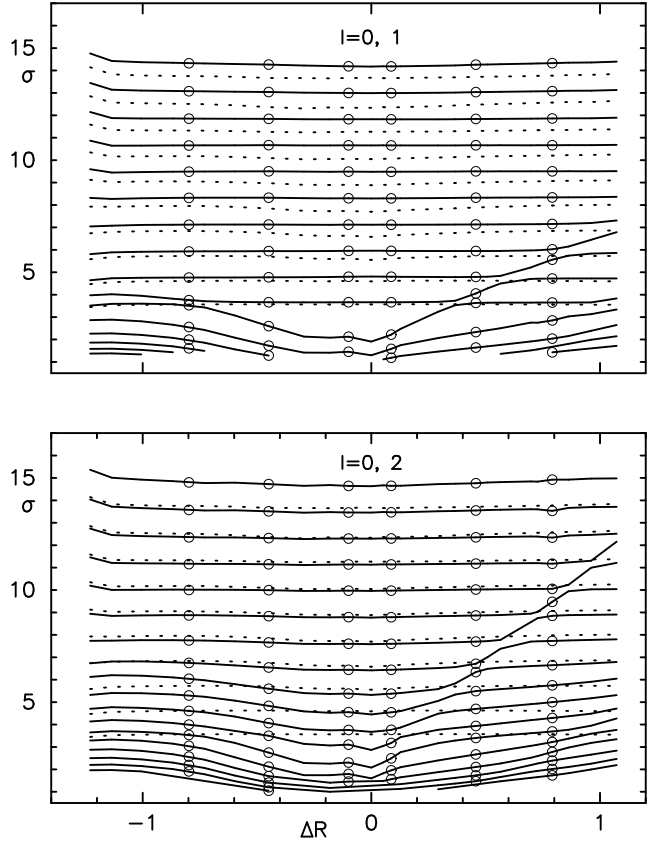}
\caption[] { Normalized frequencies of a 1.8 $M_\odot$ model 
evolving from
stage 1 to stage 25 of Fig.1. Abcissa is 
 $R-R_{ref}$ for postZams models and  $R_{ref}-R$ for pms models 
(see text). 
For sake of clarity,  
 $\ell=1$  et $\ell=2$  are plotted separately: top frame $\ell=0,1$
 and bottom frame $\ell=0,2$.
 Frequencies of radial modes 
with same radial order are connected with dotted lines; 
$\ell=1$ and   $\ell=2$  are connected  with solid lines. 
Frequencies of associated pms and
postZams models M4-M23; M8-M19; M13-M16 are presented with open circles.}
\end{figure}

Normalized
frequencies evolve  with age  quasi-symetrically 
with respect to the Zams stage. Frequencies of radial modes and nonradial
pure $p$-modes
 with given $n,\ell$  remain nearly constant when evolving from late pms to
early postZams stages because the models change nearly homologously.

On the other hand, $g$-modes, as they are sensitive to the structure of the
inner convective layers, do not behave homologously.
One can see in Fig.3 changes of 
 the frequencies of the g-modes
 which arise from the `breathing'  of the core,
(which moves back and forth)
when a new nuclear reaction starts, at models M4 and M12:
frequencies of $g$-modes
increase with age for  postZams models  because the inner maximum of 
the Br\"unt-V\"aissala frequency increases with age and
its sharp inner edge moves outward. 
 In a symmetric way with respect to the
Zams model, the frequencies of $g$-modes decrease when the pms 
model evolves toward the Zams stage 
because  the position of the inner maximum of $N$ moves inward 
with time. 

In addition, the well known phenomenon of
avoided crossing  (Unno {\sl et al.}, 1989) occurs 
during  the latest postZams
stages. The development of a $\mu$ gradient at the edge of 
the convective core of the postZams model 
causes  the development of
a sharp peak of the Br\"unt-Vaiss\"al\"a frequency. Then 
nonradial low frequency modes  undergo 
an avoided crossing which shifts the 
frequency with respect to its value in 
absence of avoided crossing.
Only those modes which are sensitive to the edge of the convective core
 have a  different
behavior before and after the Zams stage.
These interacting modes exchange their physical nature. Their frequencies
penetrate into  the $p$ mode frequency domain, breaking
the regularity seen at high frequencies.
 The progression of this phenomenon into the $p$-mode domain    
can be followed from mode with  $n \sim 1$ ($\Delta R \sim 0$)
up to $n=8$  ($\Delta R \sim 1$) for $\ell=2$ modes.
This leads to different and recognizable patterns in a
frequency spectrum, which can be considered as the signature of
different structures of the core and can
therefore tell  whether the star is a pms or a main sequence one. 

\begin{figure}
\includegraphics[width=9cm]{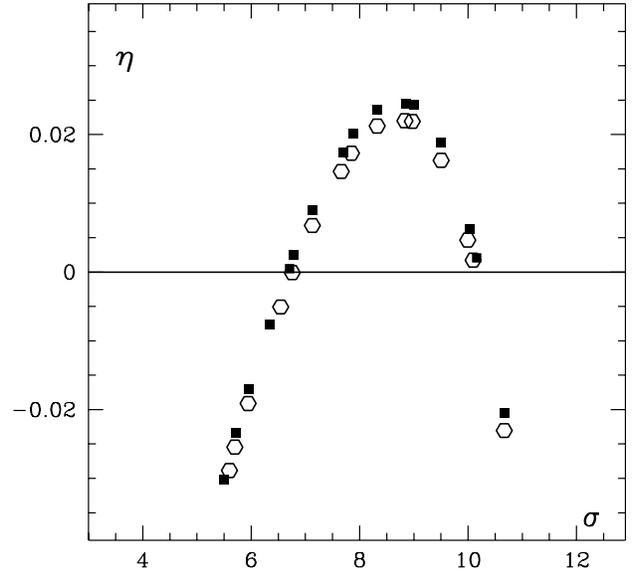}
\caption[]{Growth rate $\eta$ versus normalized frequency $\sigma$
 for low radial order $\ell=0,1,2$ modes for the associated 
 pms (M8, open circles) and main sequence  (M19, full squares)
models.}
\end{figure}
Finally, frequency calculations show the existence 
of strange modes {\sl i.e.} surface modes trapped in  very superficial outer 
layers 
(Unno {\sl et al.} 1989; Buchler {\sl et al.} 1997; Saio {\sl et al.} 1998). 
Whether these modes are excited or not  is probably strongly
dependent on the convection-pulsation interaction not included here.
As  these surface modes are not essential for our purpose, they were
disregarded in Fig.3.

\subsection {Vibrational instability}

The linear nonadiabatic calculations 
for modes with  $\ell=0,1,2$ 
for models of Table 1 show that the pms 
models lying in the classical instability strip are vibrationally
unstable. The unstable modes are the same than those for  
classical  $\delta$ Scuti stars, namely low radial order 
$g$- and $p$-modes. 

Fig.4 shows the behavior of the parameter $\eta$ (Eq.2) with increasing
normalized frequencies $\sigma$  for $\ell=0,1,2$ modes 
for the M8 pms and  the M19  main sequence  models.  The
instability strength as measured with $\eta$
 is independent of the degree $\ell$. 
 The $\eta$ curves for pms and main sequence associated models 
are almost identical. For these models, modes with radial order in the range 
 $n=1-7$ are found to be unstable.

The work integrands are found to be quite similar  for associated
pms and postZams models. Computation of the work integrals  
shows that, for both pms and
postZams models, the modes are driven  by   opacity 
 mechanisms  which efficiently operate  in the  {\it H} and{\it \ He}
ionisation zones. The  $\epsilon $ mechanism contributes only very weakly 
in the deep interior even in the youngest models.
\begin{figure}
\includegraphics[width=8.5cm]{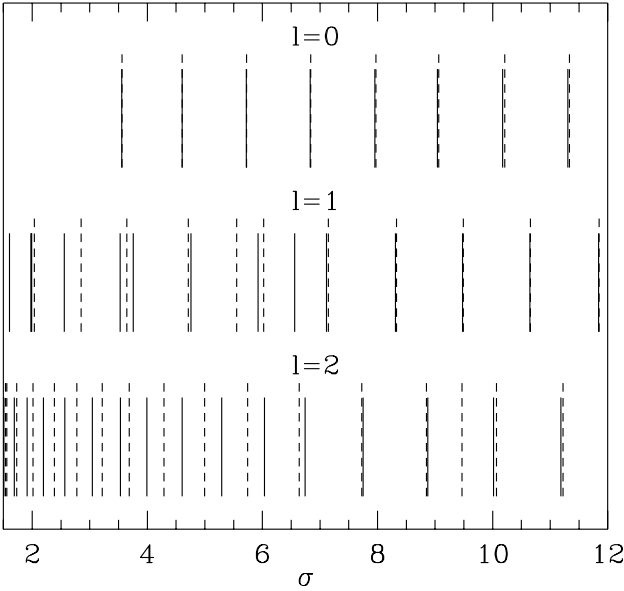}
\caption[]{ Frequency distribution for $\ell=0,1,2$ modes 
 for  pms M4 (solid lines) and main sequence M23 (dashed lines) models.  
Ordinate is  arbitrary. Abscissa is the normalized frequency $\sigma$.}
\end{figure}
\begin{figure}
\includegraphics[width=8.5cm]{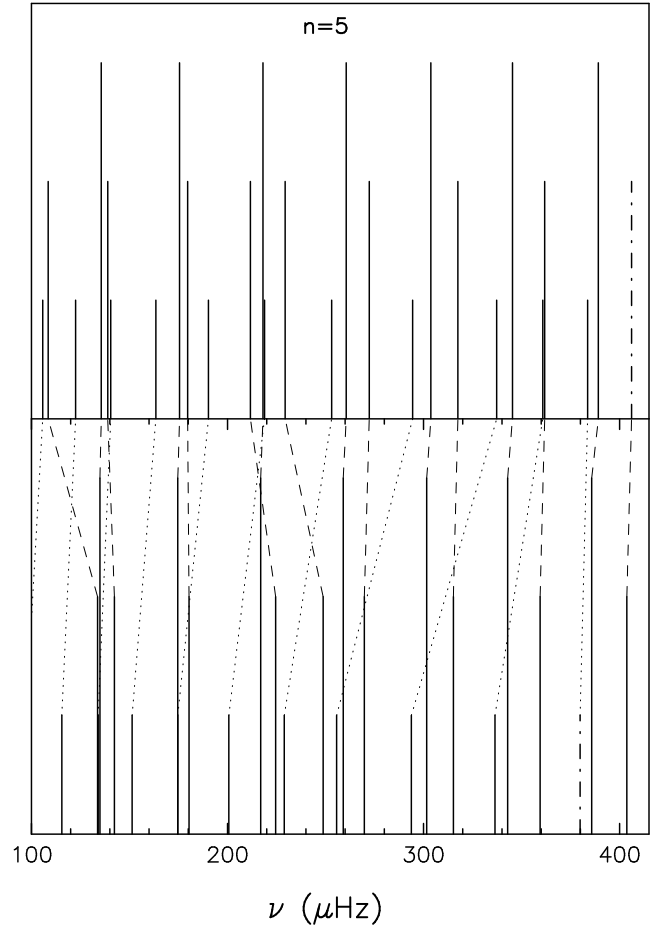}
\caption[] {~ Theoretical power spectra 
in the low frequency range where modes are excited. {\bf top:} postZams M23 
model. {\bf bottom:} pms M4 model.
 Modes with same $(\ell,n)$ subscripts in both models are 
connected with dashed lines ($\ell=0, \ell=1$ modes) 
and dotted lines ($\ell=2$) modes.
 Amplitudes are arbitrary.
 For clarity the 
highest amplitudes correspond to $\ell=0$ modes (thicker lines), 
medium ones to $\ell=1$ modes and  smallest ones to $\ell=2$ modes. 
Marginally stable modes (modes with $ -0.005 \leq \eta \leq 0 $) 
are pictured with dash-dotted vertical lines.
}
\end{figure}
 
\section{Frequency distributions of pms versus postZams models}
 Fig.5 compares frequencies of modes with the same  degree $\ell$ for the
associated models M4 and M23, represented 
respectively by solid vertical lines and dashed vertical lines. 
Frequencies of $\ell=0$ modes with same radial order are nearly identical 
as shown by the solid lines which  nearly  coincide 
with the dashed lines.
This  is because both models have similar
mean density and outer layers. This is also true for the highest frequencies 
of $\ell=1$ and $\ell=2$ modes which are associated with pure $p$ modes.  
Differences in the
spectral patterns on the other hand exist in the low frequency range.
In particular a few extra frequencies, associated with 
$\ell=1,2$ modes in avoided crossing,  exist for  the postZams  model.
As a consequence, even without a mode identification, 
a comparison of the spectra for a pms and a postZams  models at the same location
in the HR diagram  tells us whether the star is 
in a pms or a postZams stage and, in the latter
case,  yields  an indication of its age. 
\begin{figure}
\includegraphics[width=9cm]{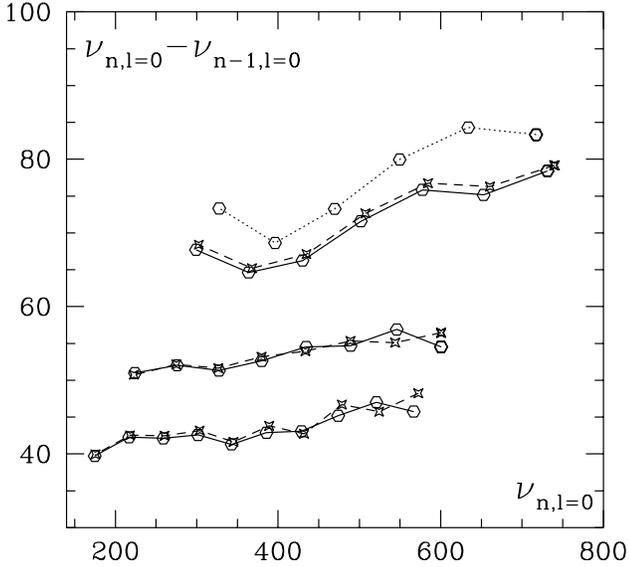}
\caption[]{ Large separations $\nu_{n \ell}-\nu_{n-1, \ell}$ ($\mu Hz$)
for low radial order $\ell = 0$ modes for 
 associated models from bottom to top 
 M4-M23; M8-M19; M13-M16 (pms:solid lines; postZams dashed lines) 
and the Zams model M15 (dotted line): large separations are the same for 
pms and postZams models with the same mean density.}
\end{figure}

Frequency differences for a given mode 
between   models M4 and M23 
 reach values as large as 50 $\mu$Hz in real units. 
This is  emphasized in Fig.6.  Global parameters are 
identical, therefore  radial modes with same radial order  
have about the  same frequencies. The external layers are 
similar, therefore the excitation range is about the same.
In this  frequency range, 
$\ell=2$ modes are the most affected by the structure of the 
deep layers and are the main agents for  differences in power spectra.

\section{Toward a seismological diagnostic: frequency differences}

Pms and postZams models at the same location in the HR
diagram have similar mean densities, 
hence one expects the large separation 
$\nu_{n,\ell} - \nu_{n-1,\ell}$ to be the same for `associated' models.
This is confirmed in Fig.7 where the large
separations for $\ell = 0$ modes (for low radial order  modes)
  are displayed for associated models 
and for the Zams
model.  The large separation is indeed the same respectively  for M4 and M23;
 for M8 and M19; for M13 and M16. The large separation scales as 
$(GM/R^3)^{1/2}$.  Given the mass, 
the large separation increases with age for
  pms models (decreasing radii) and decreases with age  
for postZams models (increasing radii); it is largest for the Zams 
model which has the smallest radius.

The development of a $\mu$ gradient at the edge of the 
convective core  can be seen on the frequency 
large separation for nonradial modes in Fig.8. 
The large separation for  $\ell=2$ modes for   associated
 pms  and  postZams  models 
 is  expressed  in normalized units 
 and  plotted against the normalized 
frequency $\sigma_{n,\ell=2}$. 
 The existence of a mode in avoided crossing
generates a large departure of the frequency difference
$\sigma_{n,\ell}-\sigma_{n-1,\ell}$ 
from its  mean value at the frequency of this mode. 
 The frequency at which a given 
mixed  mode occurs is correlated with the maximum value of the inner peak of
the Br\"unt-Vaiss\"al\"a frequency  arising from a larger $\mu$ gradient.
Such a mixed mode is detected in postZams model M20, M22, M23 and M24 in 
Fig.8.
While the main sequence model ages, the progression of the 
 mixed mode toward  higher frequencies
is clearly seen. The frequency 
of the extra mode with   the highest frequency  
 gives an indication of the age of
the postZams star.
On the other hand, pms models show a 
quasi constant large separation (in normalized units) with frequency.

Differences in the inner layers of pms and postZams models can also be  
detected by investigating the behavior of the 
small separations $\nu_{n,\ell}-\nu_{n',\ell-2}$ as a function of 
the frequency $\nu_{n,\ell}$. 
In the asymptotic regime i.e. for high radial order modes (solar like
oscillations) the small separation is defined as 
$\nu_{n,\ell}-\nu_{n-1,\ell-2}$. In the low radial order range,  avoided crossings
perturb the regular sequence of alternations of  $\ell=2$ and $\ell=0$ modes. 
 We therefore find more convenient to define  a small separation  as
the difference between a $\ell=2$ mode (radial order $n'$)
 and its closest $\ell=0$ neighbor (radial order $n$ which in some cases
at low frequency differs from $n'+1$).
Such small separations are plotted in Fig.9. When the associated models are
very close to the Zams as for instance models M13-M16,
 no difference exists between the small separations of the 
 models. For these models, as no $\mu$ gradient has developed yet,
 there is no avoided crossing  in the excited  frequency range.

\begin{figure}
\includegraphics[width=9cm]{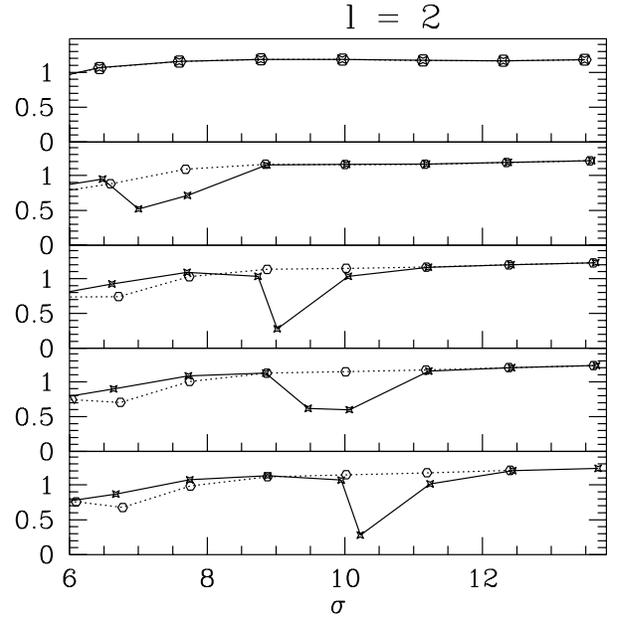}
\caption[]{ Large separations  
$\sigma_{n,\ell}- \sigma_{n-1,\ell}$ for $\ell=2$
versus $\sigma_{n,\ell=2}$
for associated pms (dotted lines) and postZams (solid lines) 
models from top to bottom: M13-M16; M7-M20; M5-M22; M4-M23; M3-M24. 
}
\end{figure}

On the other
hand,
for associated models located further away from the Zams,
for instance M4-M23, the structures of the 
inner layers are significantly different; 
this leads to a different behavior of
the  small separations of the two models 
at low frequency. 
An avoided crossing is seen to occur for modes 
with $\nu \sim 350 \mu$Hz ($\sigma \sim 8$)  
for model M23 which is not present for its associated pms model M4 
(Fig.9). At lower frequencies (below $250  \mu$Hz corresponding to
$\sigma \sim 6$), a second feature is seen
 for both models M4 and M23: this 
sharp variation of the small separation is 
due to  modes 
with frequencies close to  the value of the secondary inner
maximum of the Br\"unt-Vaiss\"al\"a 
frequency (the only inner maximum for pms models)
near the  edge of the convective core.

\section{Discussion}

 Pre-main sequence models of 1.7-2 M$_{\odot}$
crossing the classical instability strip on their way to the Zams, are
found to be vibrationally unstable, destabilized by the same mechanisms than
 the $\delta $ Scuti stars, meant as  main sequence or postZams stars.

For a given mass and a given location in the HR diagram, 
the frequencies of radial modes with same radial order  
remain nearly
 identical. As the outer layers in both stages are almost the 
same,  oscillations in
pms stars should have the same level of amplitudes as 
in  $\delta $ Scuti stars; this means that 
some modes are expected to be detectable from ground.

\begin{figure}
\vskip 0.5 truecm
\includegraphics[width=9cm]{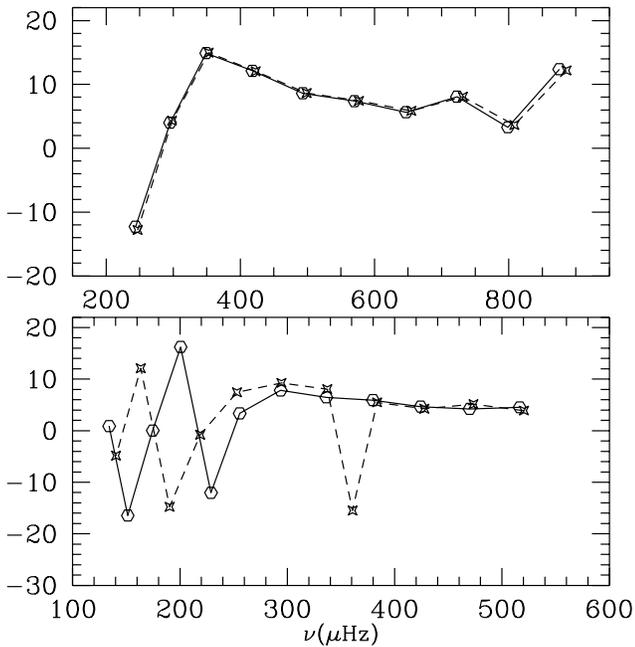}
\caption[]{ Small separations  $\nu_{n,\ell=0}- \nu_{n',\ell=2}$
versus   $\nu_{n,\ell=2}$  (in $\mu Hz$) for 
associated pms (solid lines, open circles)  and postZams
 (dashed lines, open  stars) models: M4-M23 (bottom) 
and M13-M16 (top). The  frequency difference is calculated  
between a $\ell=2$ mode and the $\ell=0$ mode  
with the closest frequency. 
}
\end{figure}

Low radial order  modes with dual (gravity and pressure) nature
penetrate in the upper layers of the
star and there is a good chance to observe them. We 
have shown that they can act  as discriminators,  as they are very
sensitive to the structure of the inner layers, 
 which differ between pms and postZams stars. 
This is particularly interesting for this region of the
 HR diagram where classical
indicators of youth  as accretion disks or 
remaining nebulosities have disappeared.

Frequency shifts and differences in the  frequency distributions
of pms and main sequence oscillators can be detected if a
sufficient number of modes are observed 
at low frequencies in the vinicity of the
fundamental radial mode. 

Important progress has been made to detect 
larger sets of frequencies from the ground using 
networks of telescopes covering all longitudes. Nevertheless 
 ideal conditions will only be
achieved in space. 
The COROT experiment (Baglin {\sl et al.}
\cite{ba}), to be launched in 2004, will 
 continuously observe  several  stars during  20 to 150
days; this will provide an  increase of a thousand for 
the signal to noise ratio and a frequency resolution from 0.5 $\mu Hz$ down
to 0.1 $\mu Hz$ .
 The large number of modes 
which is expected to be detected will help to provide, at least, a 
partial mode identification. Frequency histograms 
for instance can provide the mean density 
(Breger {\sl et al.}, 1999; Baglin {\sl et al.}, 2000; 
Barban  {\sl et al.} 2001) and the structure of the frequency spectrum
will tell whether the star is a pms or a postZams star. 

Pms stars are expected to be fast rotators 
and it will be necessary to include 
effects of rotation on the frequencies  in the same way
as proposed by Soufi {\sl et al.} (\cite{rot2}) for fast rotating $\delta$
 Scuti stars.  This  should enable  to put  constraints 
on the internal rotation of these stars (Goupil {\sl et al.}, \cite{rot1}). 
It will then be possible to trace the history- and to understand better the
process -of transfer of angular momentum inside stars with $\sim  2 M_\odot$ 
 in their early  stages of evolution.

\begin{acknowledgements}
 We are grateful to P.
Morel for making available the CESAM code.
 M.D. Suran  acknowledges  
Paris-Meudon Observatory for providing him a grant. We thank an anonymous 
referee for his comments which helped to improve the writing of this
 manuscript.
\end{acknowledgements}

\end{document}